\documentclass[12pt]{article}

\usepackage{amsmath}
\usepackage{amssymb}
\usepackage{epsfig}
\usepackage{cite}
\usepackage{color,colordvi}
\usepackage{xcolor}
\usepackage{mathtools}
\usepackage[colorlinks=false]{hyperref}

\newcommand{\kommentar}[1]{}
\newcommand{\mathd}{\mathrm{d}}
\DeclarePairedDelimiter\abs{\lvert}{\rvert}%
\DeclarePairedDelimiter\norm{\lVert}{\rVert}%

\makeatletter
\let\oldabs\abs
\def\abs{\@ifstar{\oldabs}{\oldabs*}}
\let\oldnorm\norm
\def\norm{\@ifstar{\oldnorm}{\oldnorm*}}
\makeatother

\begin{document}

\vspace{0.01cm}
\begin{center}
{\Large\bf  Damping self-forces and Asymptotic Symmetries } 

\end{center}

\vspace{0.1cm}

\begin{center}

{\bf Cesar Gomez}$^{a,b}$\footnote{cesar.gomez@uam.es}, {\bf Raoul Letschka}$^{a}$\footnote{raoul.letschka@csic.es}

\vspace{.6truecm}


{\em $^a$Instituto de F\'{\i}sica Te\'orica UAM-CSIC \\
Universidad Aut\'onoma de Madrid,
Cantoblanco, 28049 Madrid, Spain}


%


{\em $^b$Arnold Sommerfeld Center for Theoretical Physics\\
Department f\"ur Physik, Ludwig-Maximilians-Universit\"at M\"unchen\\
Theresienstr.~37, 80333 M\"unchen, Germany}

\end{center}

\begin{abstract}
\noindent
Energy conservation in radiating processes requires, at the classical level, to take into account damping forces on the sources. These forces can be represented in terms of asymptotic data and lead to charges defined as integrals over the asymptotic boundary. For scattering processes these charges, in case of zero radiated energy, are conserved and encode the information about the sub-leading soft theorems and matching conditions. The QED version of the self forces is associated with the dependence of the differential cross section on the infrared resolution scale. 
 
\end{abstract}

\newpage

In essence the information paradox \cite{Haw} reduces to identify if back reaction effects for black hole radiation are or not able to purify the final evaporation state. There is a simpler  case where to start thinking about this issue, namely considering the back reaction to electromagnetic radiation by an accelerated charged source. This is an old and extensively discussed subject where classically appears the notion of {\it self forces}. These self forces shed relevant light \cite{us} on the recently discussed subject of new asymptotic symmetries in QED \cite{Stro,Kapec:2015ena} and on the role of asymptotic dynamics. In particular the damping forces involved in radiation are intimately connected with the matching conditions as well as with the sub-leading soft theorems. Moreover the study of these damping forces, in the case of gravity, can also indicate some fruitful directions to address the most complicated case of black hole purification and the interplay between supertranslations and superrotations.

It is a well stablished fact that an accelerated charged particle loses energy. This loss can be interpreted as being caused by the action of a damping self force which in the limit of slow moving sources is given by:
\begin{equation}\label{one}
F_{d} = \frac{2 e^2 \dot a}{3 c^3}
\end{equation}
This damping self force is the radiative reaction. In other words the associated work defined by this force acting on the charge is, in time average, equal to the energy radiated by the accelerated particle. 

The simplest way to derive this damping force is expanding the retarded field created by the source ${\bf A} = \frac{1}{c} \int \frac {{\bf j}(t- \frac{r}{c})}{r} \mathd V$ at {\it short distances} i.e. in powers of $r/c$. At second order in the expansion we get
\begin{equation}\label{two}
{\bf A}^{(2)} = - \frac{2 e a}{3c^2}
\end{equation}
which leads to an electric field ${\bf E} = \frac{2 e \dot a}{3 c^3}$ that reproduces the damping force \eqref{one}. For reasons that will become clear in a moment we shall denote this field {\it Dirac field} i.e. ${\bf A}_{d}$. It is important to note that this Dirac field near the source is non singular at the position of the source. Moreover this field at time $t$ depends on the acceleration at that time.

The damping self force leads to a series of conceptual problems. In particular the equation of motion once the self force is taken into account is third order in time and has unphysical solutions where the acceleration grows exponentially in time (see \cite{Wald1} for a recent discussion). Among other reasons these problems appear due to the point like approximation for the source that is classically represented by a delta function. 

In the early days of QED it was an interesting problem to understand the meaning of this damping force. Dirac \cite{Dirac} noticed that the damping force can be written in the following way
\begin{equation}\label{abs}
F_{d} = \frac{1}{2}(F_\text{ret} -F_\text{adv})
\end{equation}
where $F_\text{ret}$ and $F_\text{adv}$ are the retarded and advanced fields created for the accelerated charge at its position. Note that this quantity is not divergent at the position of the source since the singularities of both fields cancel out. 

This comment by Dirac triggered Wheeler and Feynman to try to understand the phenomena of radiation and the radiative reaction from the absorber point of view \cite{WF}. The WF theory was based on an interesting epistemological dogma, namely an {\it isolated} system, irrespectively of its state of motion, is not radiating at all. Radiation can only happens when there is an extra charged system in the Universe able to absorb the emitted radiation (the absorber). In this picture the damping self force is interpreted not as a self force but instead as the force exerted by the absorber on the source.  The WF approach requires to modify the definition of the field created by a source to be
\begin{equation}\label{four}
\frac{1}{2}(F_\text{ret} + F_\text{adv})
\end{equation}
and to add to this field the one created by the absorber i.e. the damping self force \eqref{abs} in order to get the standard retarded field.
In the WF theory the damping self force is derived imagining that at large distances from the source you have the absorber system defined as a random distribution of charges on the asymptotic sphere. Now the action of the absorber on the radiating source can be obtained computing the {\it advanced} field created by the absorber that is perturbed by the retarded field of the source. This leads to an action of the absorber on the source at precisely the time of acceleration that reproduces in the limit of slow moving sources the damping self force. 

Independently of this absorber derivation of the damping self force, the result of Dirac allows us to write the electromagnetic field at short distances \eqref{two} in terms of {\it asymptotic data} at large distances from the source, namely
 \begin{equation}\label{Dic}
 F^{\mu\nu}_{d}(t) = \lim_{r\rightarrow\infty} r \int_{S} \frac{\mathd F_\text{ret}^{\mu\nu} (r, t-\frac{r}{c})}{\mathd t} \mathd \Omega
 \end{equation}
 where $F_\text{ret}$ is the retarded field created by the source,  $S$ represents the sphere at infinity and where $F^{\mu\nu}_{d}(t)$ is the damping field at the position of the source at time $t$. In this asymptotic limit the component that goes like $1/r^2$ is negligible. Note that asymptotically the retarded field in the integration in \eqref{Dic} is evaluated on ${\cal J}^{+}$ in case we want to use Penrose description of Minkowski. After performing the integral we can write the former expression in terms of source data as follows
 \begin{equation}\label{FDirac}
 F_{d}^{\mu\nu} = \frac{2e}{3 c^3} ( v^{\mu} \dot a^{\nu} - \dot a^{\mu} v^{\nu} )
 \end{equation}
 or equivalently as $\frac{d}{dt}(v^{\mu} a^{\nu} -a^{\mu} v^{\nu}) $.
We can denote this quantity as ${\cal Q}^{+}(t)$. Thus we can define
 \begin{equation}
 Q^{+} = \int \mathd t {\cal Q}^{+}(t)
 \end{equation}
 that measures the total amount of radiated energy. Note that this integral can be written as an integral over the retarded time coordinate $u$ of  ${\cal J}^{+}$ and in that sense the former charge is defined as an integral over the asymptotic boundary ${\cal J}^{+}$.
 
 The soft part of ${\cal Q}^{+}$ leads to a part of $Q^{+}$ that only depends on the scattering data and that contains the information about the {\it sub-leading soft photon theorems} \cite{sub, Conde:2016csj,Campiglia:2014yka,diVecchia}. Indeed we can think of charges of type $\int_{S} F \varepsilon$ where $\varepsilon$ is a function on the sphere that enters as a convolution and {\it damping charges}
 \begin{equation}\label{Dampingcharge}
 \int_{S} r\frac{\mathd F^{\mu\nu}}{\mathd t} \hat u_{\nu} \varepsilon_{\mu}
 \end{equation}
 for $\hat u$ the unit velocity vector and $\varepsilon_{\mu} = \partial_{\mu}\varepsilon$. In this case the charges are informing us about the radiated energy. We can also consider the radiated angular momentum $ \int_{S} r\frac{\mathd F^{\mu\nu}}{\mathd t} \hat u_{\mu} \times \bf r$. Schematically:
 
 \vspace{5mm}
 
 soft theorem $\rightarrow$ electrostatic dressing of asymptotic states\footnote{The FK $S$-matrix \cite{FK} only accounts for this asymptotic electrostatic dressing. See also references \cite{porrati,amit,panchenko,stromingernew,panchenko2}}.
 
 \vspace{5mm}
 
 sub leading soft theorems $\rightarrow$ damping self forces at the interaction region.
 
 \vspace{5mm}
 In the electrostatic dressing case the underlying symmetry is simply gauge invariance, something that is classically clear since the electrostatic dressing only depends on Gauss law. In the the case of damping to unveil, from the classical Lagrangian point of view, the underlying symmetry, is a bit more subtle. The reason is that damping self forces cannot be implemented in a Lagrangian formalism, since, as already pointed out, lead to equations of motion which are third order.

  Using the time reversal version of the damping self force 
  \begin{equation}
  \lim _{r\rightarrow \infty}\int_{S} r \frac{\mathd{ F_\text{adv}}}{\mathd t} \mathd \Omega
  \end{equation}
  we can define analog charges ${\cal Q}^{-}(t)$. The integrand now can be view as living on $\cal J^{-}$ and the charge $Q^{-}$ as integrals over ${\cal J}^{-}$ \footnote{Where ${\cal J}^{-}$ has the topology $R\times S$.}. The conservation law for the corresponding charges, for instance in a scattering process, 
 \begin{equation} \label{conserv}
  { Q}^{+}={ Q}^{-}
  \end{equation}
  simply reflects the condition that the incoming radiated energy is equal to the outgoing radiated energy and it corresponds to the {\it no net radiation} matching condition $F_\text{ret}=F_\text{adv}$. The soft theorems account for the zero mode version of this conservation law.  If we think in a generic scattering process among electrons we can impose this condition if we reduce to processes where the incoming energy carried by incoming soft photons is equal to the outgoing radiated energy. We will come back briefly to this point in a moment.

 The important lesson that can be derived from the former discussion is that for a radiating source the {\it short distance} behaviour of the Dirac field describing the damping is determined by the {\it asymptotic dynamics}. To implement this damping effect, that is needed in order to achieve conservation of energy, makes the scattering process asymmetric in time.  Time symmetry at the level of the $S$-matrix can be only achieved if the stronger condition of zero net radiated energy is imposed or equivalently matching conditions leading to the conservation laws \eqref{conserv}.
 
  Let us briefly discuss the time asymmetry of radiation. One of the original targets of WF discussion was to understand the {\it irreversibility} of radiation processes within a theory as classical electrodynamics that is perfectly time symmetric. The irreversibility intrinsic to radiation is understood in WF theory as statistical in origin (see \cite{Einstein} for the historical origin of the discussion). The reason can be easily understood in standard QED. In the radiation process the incoming energy is {\it ordered} while the damping force on the sources needed to achieve energy conservation is accounted in the cross section by a statistical ensemble of unresolved soft photons carrying total energy equal to the one involved in the damping. Note that in QED this irreversibility is not in the $S$-matrix that is perfectly time symmetric, but is statistical in origin and shows up in the differential cross section where the phase space factors for the soft radiation are large \cite{us}. This can be also understood recalling the role of the soft photon theorem in the infrared divergences of QED. The soft photon theorem 
  \begin{equation}
  \lim _{k\rightarrow0}A(p_i,p_f,k) = \lim_{k\rightarrow0}S(k;p_i,p_f) A(p_i,p_f)
  \end{equation}
where we denote $S$ the soft photon pre factor, can be {\it exponentiated} if we consider the sum $\sum_n \frac{1}{n!}A(p_i,p_f,k,k,..)$ for emission of $n$ soft photons. This sum leads for the cross section the well known factor $e^{\int \mathd k S(p_i,p_f;k)}$ where is this integral over $k$ coming from the phase space integral where we need to include the infrared resolution scale $\varepsilon = \sum k$. In other words {\it the exponentiation} of the would be generator of the symmetry underlying the soft photon theorem requires, in order to be defined, to work with the cross section and to include the infrared scale.
 
 Can the former discussion have some applications to gravity and black hole physics? The answer requires to define the analog of the damping Dirac self field for gravity and incidentally also for the black hole radiation process. In linearised gravity the analog of the Dirac damping self field has been extensively discussed (see for instance \cite{Wald2,Wald3}). Likely, in the language of BMS asymptotic symmetries, {\it the damping reaction} to supertranslation projects non trivially on  superrotations. For black hole radiation the corresponding Dirac field could be thought as living on the horizon (or on the stretched horizon) reflecting, the damping effect on the horizon, of the radiated energy\footnote{This Dirac field can be possibly thought in terms of some sort of super translation field on the horizon \cite{HPS} or $A$ field \cite{Artem}.}. A rough estimate indicates that this damping self force is $O(1/N)$ for $N= M^2L_P^2$ and $M$ the black hole mass\footnote{This would be in agreement with the expected back reaction purifying effects \cite{dvali}}. The whole process of creation and evaporation is expected to be perfectly unitary although, as it is the case in QED, the damping back reaction effects need to be described {\it statistically } at the level of the cross section.
 
 \section*{Acknowledgements}
 The material of this note was presented by CG in the workshop ``New Results in Quantum Field Theory and Holography'' at Trinity College in Dublin. I would like to thank the organizers for the great and stimulating atmosphere.

The work of C.G. was supported in part by Humboldt Foundation and by Grants: FPA 2009-07908 and  ERC Advanced Grant 339169 "Selfcompletion''. The work of R.L was supported by the ERC Advanced Grant 339169 "Selfcompletion''.

\end{document}